\documentclass[12pt,preprint]{aastex}

\usepackage{rotating}

\newcommand{\pivec}{\mbox{\boldmath $\pi$}}

\lefthead{HAN ET AL.} 
\righthead{STRONG LENSING PERTURBATION AT MODERATE MAGNIFICATION}

\begin{document}
\title{Interpretation of Strong Short-Term Central Perturbations in the Light
Curves of Moderate-Magnification Microlensing Events}

\author{
C. Han\altaffilmark{1,2},
K.-H. Hwang\altaffilmark{2},
D. Kim\altaffilmark{2},
A. Udalski\altaffilmark{3,4},
F. Abe\altaffilmark{9,10},
L.A.B. Monard\altaffilmark{1,5},
J. McCormick\altaffilmark{1,18}\\
and\\
M.K. Szyma\'nski\altaffilmark{4},
M. Kubiak\altaffilmark{4},
G. Pietrzy\'nski\altaffilmark{4,14},
I. Soszy\'nski\altaffilmark{4},
O. Szewczyk\altaffilmark{14},
\L. Wyrzykowski\altaffilmark{13},
K. Ulaczyk\altaffilmark{4}\\
(The OGLE Collaboration),\\
I.A. Bond\altaffilmark{25},
C.S. Botzler\altaffilmark{26},
A. Fukui\altaffilmark{8},
K. Furusawa\altaffilmark{8},
J.B. Hearnshaw\altaffilmark{27},
Y. Itow\altaffilmark{8},
K. Kamiya\altaffilmark{8},
P.M. Kilmartin\altaffilmark{28},
A. Korpela\altaffilmark{29},
W. Lin\altaffilmark{25},
C.H. Ling\altaffilmark{25},
K. Masuda\altaffilmark{8},
Y. Matsubara\altaffilmark{8},
N. Miyake\altaffilmark{8},
Y. Muraki\altaffilmark{30},
M. Nagaya\altaffilmark{8},
K. Ohnishi\altaffilmark{31},
T. Tokumura\altaffilmark{8},
Y.C. Perrott\altaffilmark{26},
N. Rattenbury\altaffilmark{22},
To. Saito\altaffilmark{32},
T. Sako\altaffilmark{8},
L. Skuljan\altaffilmark{25},
D.S. Sullivan\altaffilmark{29},
T. Sumi\altaffilmark{8},
W.L. Sweatman\altaffilmark{25},
P.,J. Tristram\altaffilmark{28},
P.C.M. Yock\altaffilmark{26}\\
(The MOA Collaboration),\\
W. Allen\altaffilmark{13},
G.W. Christie\altaffilmark{14},
D.L. DePoy\altaffilmark{15},
S. Dong\altaffilmark{16},
B.S. Gaudi\altaffilmark{16},
A. Gould\altaffilmark{16},
C.-U. Lee\altaffilmark{17},
T. Natusch\altaffilmark{19},
B.-G. Park\altaffilmark{17},
R.W. Pogge\altaffilmark{16}\\
(The $\mu$FUN Collaboration),\\
M.D.~Albrow\altaffilmark{27},
A. Allan\altaffilmark{20},
V. Batista\altaffilmark{33},
J.P.~Beaulieu\altaffilmark{33},
D.P. Bennett\altaffilmark{10},
S.~Brillant\altaffilmark{23},
M.~Bode\altaffilmark{21},
D.M.~Bramich\altaffilmark{35},
M.~Burgdorf\altaffilmark{21},
J.A.R.~Caldwell\altaffilmark{36},
H. Calitz\altaffilmark{37},
A.~Cassan\altaffilmark{38},
E.~Corrales\altaffilmark{33},
S.~Dieters\altaffilmark{33,34},
D.~Dominis Prester\altaffilmark{39},
M.~Dominik\altaffilmark{6},
J.~Donatowicz\altaffilmark{40},
P. Fouque\altaffilmark{41},
J. Greenhill\altaffilmark{34},
K.~Hill\altaffilmark{34},
M.~Hoffman\altaffilmark{37},
K.~Horne\altaffilmark{6},
U.G.~J{\o}rgensen\altaffilmark{42},
N.~Kains\altaffilmark{6},
S.~Kane\altaffilmark{16},
S.~Kubas\altaffilmark{23},
J.B.~Marquette\altaffilmark{33},
R.~Martin\altaffilmark{43},
P.~Meintjes\altaffilmark{37},
J.~Menzies\altaffilmark{44},
K.R.~Pollard\altaffilmark{27},
K.C.~Sahu\altaffilmark{9},
C.~Snodgrass\altaffilmark{23},
I.~Steele\altaffilmark{21},
R. Street\altaffilmark{24},
Y.~Tsapras\altaffilmark{24},
J.~Wambsganss\altaffilmark{41},
A.~Williams\altaffilmark{43},
M.~Zub\altaffilmark{41}\\
(The PLANET/RoboNet Collaboration)\\
}

\altaffiltext{1}{Microlensing Follow Up Network ($\mu$FUN)}
\altaffiltext{2}{Department of Physics, Chungbuk National University, Cheongju 361-763, Korea}
\altaffiltext{3}{Optical Gravitational Lens Experiment (OGLE)}
\altaffiltext{4}{Warsaw University Observatory, Al. Ujazdowskie 4, 00-478 Warszawa, Poland}
\altaffiltext{5}{Bronberg Observatory, Centre for Backyard Astrophysics Pretoria, South Africa}
\altaffiltext{6}{SUPA, Physics \& Astronomy, North Haugh, St Andrews, KY16 9SS, UK}
\altaffiltext{7}{Microlensing Observations in Astrophysics (MOA) Collaboration}
\altaffiltext{8}{Solar-Terrestrial Environment Laboratory, Nagoya University, Nagoya, 464-8601, Japan}
\altaffiltext{9}{Space Telescope Science Institute, 3700 San Martin Drive, Baltimore, MD 21218, USA}
\altaffiltext{10}{Department of Physics, Notre Dame University, Notre Dame, IN 46556, USA}
\altaffiltext{11}{Institute of Astronomy Cambridge University, Madingley Rd., CB3 0HA Cambridge, UK}
\altaffiltext{12}{Universidad de Concepci\'on, Departamento de Fisica, Casilla 160-C, Concepci\'on, Chile}
\altaffiltext{13}{Vintage Lane Observatory, Blenheim, New Zealand}
\altaffiltext{14}{Auckland Observatory, Auckland, New Zealand}
\altaffiltext{15}{Department of Physics, Texas A\&M University, College Station, TX, USA}
\altaffiltext{16}{Department of Astronomy, Ohio State University, 140 W. 18th Ave., Columbus, OH 43210, USA}
\altaffiltext{17}{Korea Astronomy and Space Science Institute, Daejeon 305-348, Korea}
\altaffiltext{18}{Farm Cove Observatory, Centre for Backyard Astrophysics, Pakuranga, Auckland New Zealand}
\altaffiltext{19}{AUT University, Auckland, New Zealand}
\altaffiltext{20}{School of Physics, University of Exeter, Stocker Road, Exeter EX4 4QL, UK}
\altaffiltext{21}{Astrophysics Research Institute, Liverpool John Moores Univ., Twelve Quays House, Egerton Wharf, Birkenhead CH41 1LD, UK}
\altaffiltext{22}{Jodrell Bank Centre for Astrophysics, University of Manchester, Manchester, M13 9PL,UK}
\altaffiltext{23}{European Southern Observatory, Alonso de Cordova 3107, Casilla 19001, Vitacura, Santiago 19, Chile}
\altaffiltext{24}{Las Cumbres Observatory Global Telescope Network, 6740B Cortona Dr, Suite 102, Goleta, CA, 93117, USA}
\altaffiltext{25}{Institute of Information and Mathematical Sciences, Massey University, Private Bag 102-904, North Shore Mail Centre, Auckland, New Zealand}
\altaffiltext{26}{Department of Physics, University of Auckland, Private Bag 92019, Auckland, New Zealand}
\altaffiltext{27}{University of Canterbury, Department of Physics and Astronomy, Private Bag 4800, Christchurch 8020, New Zealand}
\altaffiltext{28}{Mt. John Observatory, P.O. Box 56, Lake Tekapo 8770, New Zealand}
\altaffiltext{29}{School of Chemical and Physical Sciences, Victoria University, Wellington, New Zealand}
\altaffiltext{30}{Department of Physics, Konan University, Nishiokamoto 8-9-1, Kobe 658-8501, Japan}
\altaffiltext{31}{Nagano National College of Technology, Nagano 381-8550, Japan}
\altaffiltext{32}{Tokyo Metropolitan College of Industrial Technology, Tokyo 116-8523, Japan}
\altaffiltext{33}{Institut d'Astrophysique de Paris, 98bis Boulevard Arago, 75014 Paris, France}
\altaffiltext{34}{University of Tasmania, School of Maths and Physics, Private bag 37, GPO Hobart, Tasmania 7001, Australia}
\altaffiltext{35}{Issac Newton Group, Apartado de Correos 321, E-38700 Santa Cruz de La Palma, Spain}
\altaffiltext{36}{McDonald Observatory, 16120 St Hwy Spur 78, Fort Davis, TX 79734, USA}
\altaffiltext{37}{Dept. of Physics, Boyden Observatory, University of the Free State, Bloemfontein 9300, South Africa}
\altaffiltext{38}{Observatoire Midi-Pyr\'en\'ees, UMR 5572, 14, avenue Edouard Belin, 31400 Toulouse, France}
\altaffiltext{39}{Physics department, Faculty of Arts and Sciences, University of Rijeka, 51000 Rijeka, Croatia}
\altaffiltext{40}{Technical University of Vienna, Dept. of Computing, Wiedner Hauptstrasse 10, Vienna, Austria}
\altaffiltext{41}{Astronomisches Rechen-Institut, Zentrum f\"ur Astronomie, Heidelberg University, M\"{o}nchhofstr. 12-14, 69120 Heidelberg, Germany}
\altaffiltext{42}{Niels Bohr Institute, Astronomical Observatory, Juliane Maries Vej 30, DK-2100 Copenhagen, Denmark}
\altaffiltext{43}{Perth Observatory, Walnut Road, Bickley, Perth 6076, Australia}
\altaffiltext{44}{South African Astronomical Observatory, P.O. Box 9 Observatory 7935, South Africa}


\begin{abstract}
To improve the planet detection efficiency, current planetary microlensing 
experiments are focused on high-magnification events searching for 
planetary signals near the peak of lensing light curves.
However, it is known that central perturbations can also be produced 
by binary companions and thus it is important to distinguish 
planetary signals from those induced by binary companions.
In this paper, 
we analyze the light curves of microlensing events 
OGLE-2007-BLG-137/MOA-2007-BLG-091,
OGLE-2007-BLG-355/MOA-2007-BLG-278, and
MOA-2007-BLG-199/OGLE-2007-BLG-419, for all
of which exhibit short-term perturbations near the peaks 
of the light curves.  From detailed modeling of the light curves, 
we find that the perturbations of the events are caused by binary 
companions rather than planets.  From close examination of the 
light curves combined with the underlying physical geometry of 
the lens system obtained from modeling, we find that the short 
time-scale caustic-crossing feature occurring at a low or a moderate 
base magnification with an additional secondary perturbation is a 
typical feature of binary-lens events and thus can be used for the 
discrimination between the binary and planetary interpretations.
\end{abstract}



\section{Introduction}

The microlensing signal of a planet is characterized by a short-term 
perturbation to the smooth standard single-lens light curve of the
primary-induced lensing event occurring on a background source star
\citep{mao91,gould92}.  The duration of the perturbation is several 
days for a gas giant and several hours for an Earth-mass planet, 
while the typical cadence of lensing surveys is roughly a day. As 
a result, it is difficult to detect planets from the survey observations 
alone.  To achieve the cadence required to detect short-term planetary 
signals, current planetary lensing searches are being conducted 
by combining survey and follow-up observations, in which the survey 
observations \citep{udalski03,bond02} aim to maximize the number of 
detections of lensing events by monitoring a large area of sky and 
follow-up observations \citep{albrow98,ghosh04} are focused on 
intensive monitoring of the events detected by the survey observations.  
However, the number of telescopes available for follow-up observations 
is far smaller than would be needed to monitor all the detected events 
from the survey observations.  As a result, follow-up observations are 
selectively conducted for events that can maximize the planet detection 
probability.  Currently, the highest priority is given to high-magnification 
events.  For these events, the planet detection efficiency is high 
because the source trajectory always passes close to the region of 
central perturbations.  In addition, the perturbation occurs near the 
peak of the light curve and thus follow-up observations can be prepared 
in advance.

Although the strategy to detect central perturbations has an 
important advantage of high sensitivity to planets, it also has 
a shortcoming of difficulty in the interpretation of observed 
signals.  One important cause of this difficulty is that the 
perturbation near the peak of a lensing light curve can be produced 
not only by a planet but also by a wide or a close binary companion 
with a mass roughly equal to that of the primary. Fortunately, 
the perturbation patterns produced by the planetary and the binary 
companion are intrinsically different \citep{han08, han09a}, and 
thus it is in principle possible to discriminate between the 
planetary and binary interpretations.  However, this discrimination 
usually requires detailed modeling of the light curve, which demands 
a time-consuming search for a solution of the lensing parameters in 
the vast space of many parameters.  Therefore, simple diagnostics 
that can be used for the discrimination between the two interpretations 
will be useful not only for the immediate identification of the nature 
of the perturbation but also for the preparation of observational 
strategies for better characterization of planetary systems.

In this paper, we present the results of the analysis of three 
{\it binary-lens} events with very strong short-term signals near 
the peaks of the light curves.  From the investigation of the light 
curves combined with the underlying physical geometry of the lens 
system, we find and present several features in lensing light curves  
that can be used as diagnostic tools for the discrimination between 
the planetary and binary interpretations of observed events.

The paper is organized as follows.  In section 2, we briefly describe 
the general features of central perturbations produced by a planet 
and a binary companion.  In section 3, we discuss the data used in 
our analysis.  In section 4, we explain the modeling procedure and 
present the best-fit lensing parameters of the individual lensing 
events obtained from the modeling.  In section 5, we present the 
features in the light curves that characterize the binary nature of 
the events and discuss the usability of the features in distinguishing 
the planetary and binary interpretations. We summarize the results and 
conclude in section 6.

\section{Planetary and Binary Central Caustics}

Lens systems composed of multiple masses can produce strong 
perturbations when the trajectory of a source star passes close 
to a caustic. The caustic represents the source star position
at which the magnification of a point source becomes infinite. 
For lenses consisting of two masses, the set of caustics form 
closed curves each of which is composed of concave curves that 
meet at cusps.

Binary lenses can have one, two, or three sets of caustic depending 
on the separation between the lens components.  If the two masses 
are separated by approximately an Einstein radius, then there is 
a single six-cusp caustic. If the separation is substantially 
closer than the Einstein radius (close binary), there is a single 
four-cusp central caustic and two small outlying three-cusp caustics. 
If the masses are separated by substantially more than the Einstein 
radius (wide binary), there are two four-cusp caustics each of which 
is associated with a member of the binary.  The short-term central 
perturbation of a binary-lens event can occur in two different ways.  
The first case is when the trajectory of a source star passes close 
to the central caustic of a close binary.  The other case happens 
when the trajectory passes close to one of the two caustics of a 
wide-separation binary.  For both wide and close binaries, the 
central caustics have a diamond shape although there is some variation 
depending on the separation and the mass ratio between the two binary 
components.

For planetary lensing, the central caustic refers to a small 
four-cusp caustic located close to the primary lens.  The size 
of the planetary central caustic as measured by the separation 
between the two cusps located on the primary-planet axis is 
related to the planetary lensing parameters by \citep{chung05}
\begin{equation}
\Delta\xi_{\rm c}\sim {4q\over (s-s^{-1})^2},
\label{eq1}
\end{equation}
where $s$ represents the projected primary-planet separation 
normalized by the Einstein radius corresponding to the total 
mass of the lens system, $\theta_{\rm E}$, and $q$ is the 
planet/primary mass ratio.  While the size of the central 
caustic depends on both the mass ratio and the separation, 
its shape depends solely on the separation. When the shape 
is quantified as the ratio of the vertical width 
$\Delta\eta_{\rm c}$ to the horizontal width $\Delta\xi_{\rm c}$, 
the width ratio is related to the planetary separation by 
\citep{chung05}
\begin{equation}
{\Delta\eta_{\rm c}\over\Delta\xi_{\rm c}}\sim {
(s-s^{-1})|\sin^3\phi|
\over 
(s+s^{-1}-2\cos\phi)^2
},
\label{eq2}
\end{equation}
where $\cos\phi=(3/4)(s+s^{-1})\{ 1-[(32/9)(s+s^{-1})^2]^{1/2}\}$.
For the range of the planetary separations within which the planet 
detection efficiency is not negligible, the width 
ratio of the planetary central caustic is substantially smaller 
than unity \citep{han09b}, implying that the caustic is elongated along the 
primary-planet axis.  The caustic has an arrow-head shape where 
three cusps are located close to the primary and the other cusp 
is located at the tip of the arrow pointing toward the direction 
of the planet.  For a given mass ratio, a pair of central caustics 
with separations $s$ and $s^{-1}$ are identical to the first order 
of approximation \citep{dominik99,an05}.

\section{Observations}

\begin{figure*}[t]
\epsscale{0.9}
\plotone{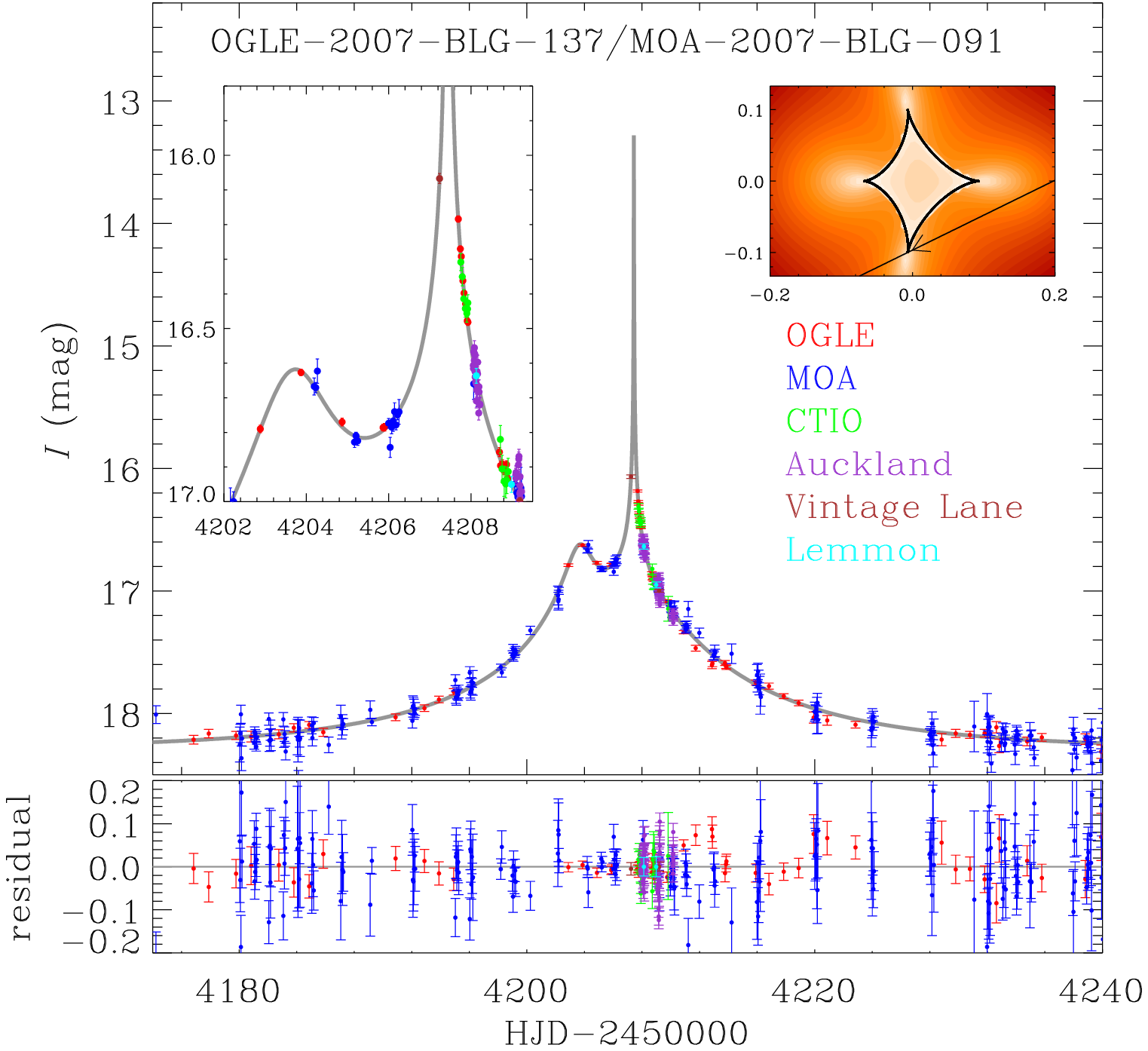}
\caption{\label{fig:one}
Data and best-fit model of the microlensing event 
OGLE-2007-BLG-137/MOA-2007-BLG-091. The left inset shows a blowup 
of the region of major perturbation.  The right inset shows the 
geometry of the source trajectory with respect to the caustic 
location.  The coordinates are centered at the `center of magnification' 
of the lens system (see the definition in the text) and the abscissa 
and the ordinate are parallel with and normal to the binary axis, 
respectively.  Heavier mass is on the left.  The temperature scale 
represents the magnification where brighter tone represents higher 
magnification. The bottom panel represents the residual from the 
model light curve.
}\end{figure*}

\begin{figure*}[t]
\epsscale{0.9}
\plotone{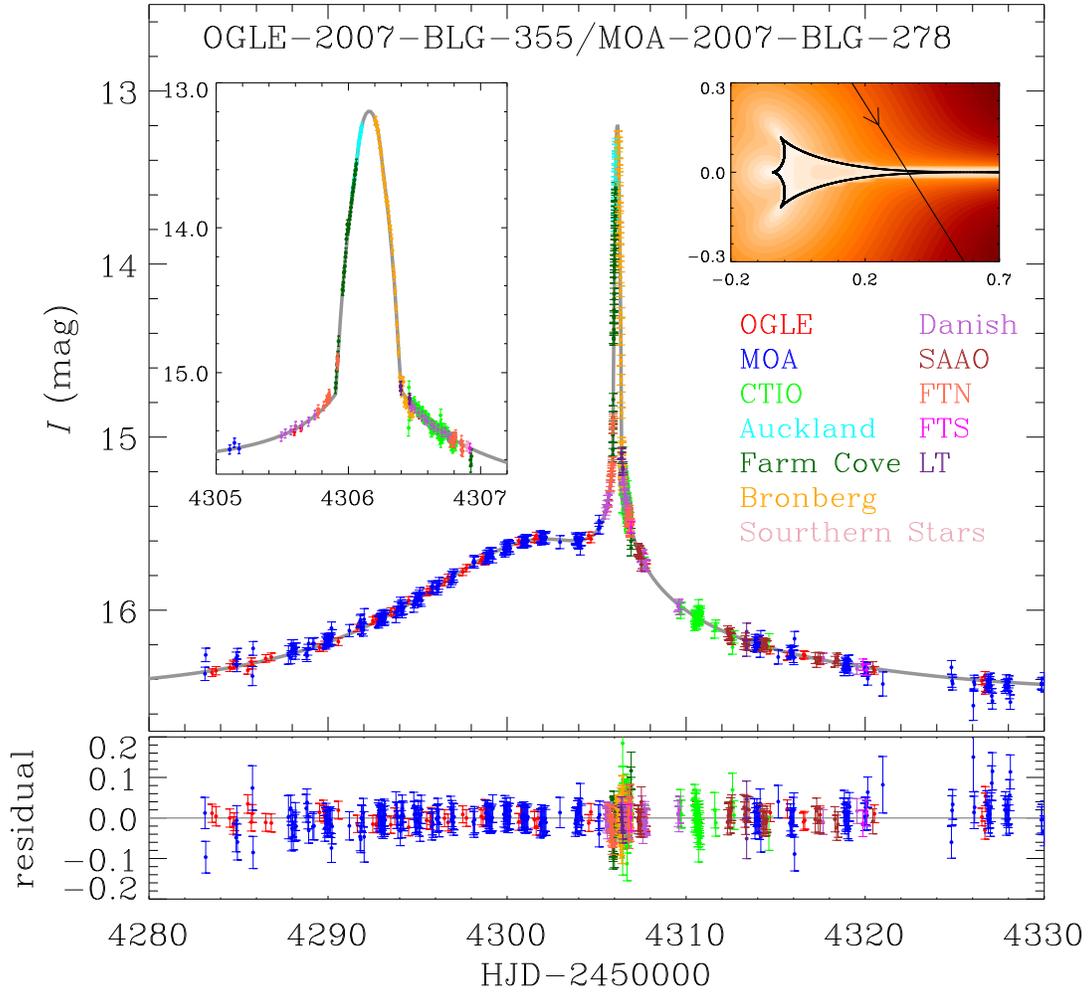}
\caption{\label{fig:two}
Data and best-fit model of OGLE-2007-BLG-355/MOA-2007-BLG-278.
Notations are same as in Fig.~\ref{fig:one}.
}\end{figure*}

\begin{figure*}[t]
\epsscale{0.9}
\plotone{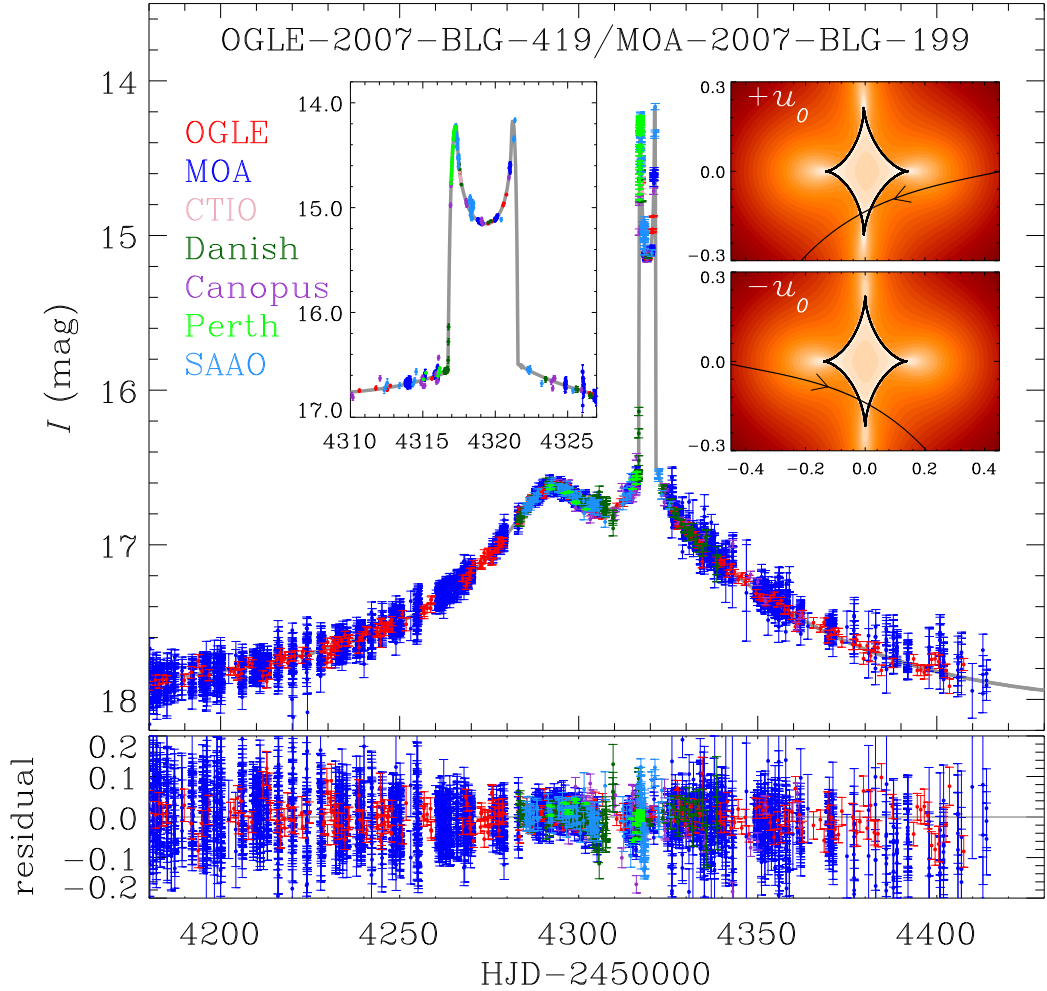}
\caption{\label{fig:three}
Data and best-fit model of MOA-2007-BLG-199/OGLE-2007-BLG-419.
Notations are same as in Fig.~\ref{fig:one} except that there 
are two insets of source trajectory geometry for the two 
degenerate models. The presented model light curve is for 
``$+u_0$'' solution (see Table \ref{table:one}), which has 
smaller $\Delta\chi^2$ by $\sim 90$.
}\end{figure*}

The events we analyze include OGLE-2007-BLG-137/MOA-2007-BLG-091,
OGLE-2007-BLG-355/MOA-2007-BLG-278, and MOA-2007-BLG-199/OGLE-2007-BLG-419.
All these events were observed in the 2007 microlensing observation 
season.  The light curves of the individual events are presented 
in Figure~\ref{fig:one} -- Figure~\ref{fig:three}.  They show a 
common feature of strong perturbations occurring near the peaks 
of the light curves.  Due to the location of the perturbations 
similar to those of central perturbations induced by planets, all 
events were alerted by survey observations and intensively monitored 
by follow-up observations.

The lensing event OGLE-2007-BLG-137/MOA-2007-BLG-091 was alerted on 
2007 April 15 by the OGLE collaboration as a possible event with an 
anomaly near the peak, using the 1.3 m Warsaw telescope of Las 
Campanas Observatory in Chile.  In response to the alert, follow-up 
observations were conducted by the $\mu$FUN team with various 
telescopes including the 1.3 m of CTIO in Chile, the 0.4 m of 
Auckland Observatory, the 0.4 m of Vintage Lane Observatory in 
New Zealand, and the 1.0 m of Mt.\ Lemmon Observatory in Arizona, 
United States.  The MOA collaboration independently discovered the 
event by using the 1.8 m of Mt.\ John in New Zealand.  From these 
observations, a total of 457, 1119, 14, 65, 2, and 2 images were 
obtained from Las Campanas, Mt.\ John, CTIO, Auckland, Vintage Lane, 
and Mt.\ Lemmon, respectively.

The event OGLE-2007-BLG-355/MOA-2007-BLG-278 was detected independently 
by the OGLE and MOA survey groups and its anomalous behavior was 
noticed by the SIGNALMEN, a software developed 
by \citet{dominik07} for early detection of anomalies and 
now integrated part of the ARTEMiS system \citep{dominik08}, on 2007 
July 20.  Soon after, the $\mu$FUN and PLANET/RoboNet teams conducted 
intensive follow-up observations by using the 1.3 m of CTIO, the 0.4 m 
of Auckland Observatory, the 0.36 m of Farm Cove Observatory in New 
Zealand, the 0.28 m of Southern Stars Observatory in Tahiti, the 0.36 m 
of Bronberg Observatory in South Africa, 1.54 m Danish Telescope of La 
Silla Observatory in Chile, the 1.0m of SAAO in South Africa, 2.0 m 
Faulkes S. (FTS) Telescope in Australia, 2.0 m Faulkes N. (FTN) Telescope 
in Hawaii, and 2.0 m Liverpool Telescope (LT) in La Palma, Spain. 
Due to the prompt response to the alert, the perturbation was very 
densely resolved by the follow-up observations despite the fact that 
the perturbation lasted only about 1 day.  The data consist of 591, 
1325, 187, 21, 39, 3, 43, 40, 34, 3, 27, and 7 images obtained from 
Las Campanas, Mt.\ John, CTIO, Auckland, Farm Cove, Southern Star, 
Bronberg, La Silla, SAAO, FTS, FTN, and LT, respectively.

The event MOA-2007-BLG-199/OGLE-2007-BLG-419 is a long time-scale 
event. The PLANET/RoboNet team first announced the anomalous behavior 
of the event on 2007 July 28 and conducted follow-up observations by 
using the 1.54 m Danish Telescope, the 1.0 m of Canopus Observatory, 
the 0.6 m of Perth Observatory in Australia, and the 1.0m of SAAO.  
The $\mu$FUN collaboration also observed the event by using the 1.3m 
of CTIO.  The event shows a clear signature of caustic crossing during 
HJD $\sim 2454316$ -- 2454322.  The data consist of 1233, 4108, 126, 
56, 52, 130, and 15 images obtained from Las Campanas, Mt.\ John 
University, La Silla, Canopus, Perth, SAAO, and CTIO, respectively.

For the analysis of the light curves, we use the photometry results 
that were reduced by the individual groups using their own photometry 
codes.

\section{Modeling}

Binary-lens modeling requires inclusion of various parameters.  
The first set of three parameters is related to the geometry 
of a single-lens event.  These include the Einstein time scale, 
$t_{\rm E}$, the time of maximum magnification, $t_0$, and the 
lens-source separation normalized by the Einstein radius 
$\theta_{\rm E}$ at the time of maximum magnification, $u_0$.  
For the description of the perturbation, an additional set of 
binary parameters is required.  These parameters include the 
mass ratio $q$ and separation $s$ between the lens components 
and the angle of the source trajectory with respect to the 
binary axis, $\alpha$.  In addition, the normalized source 
radius, $\rho_\star\equiv\theta_\star/\theta_{\rm E}$, is 
needed to describe the lensing magnification whenever the 
angular radius of the source star $\theta_\star$ plays an 
important role, such as the moment of a caustic crossing.  
For some events with long time scales, it is required to 
include the parallax parameters $\pi_{{\rm E},N}$ and 
$\pi_{{\rm E},E}$ to account for the deviation of the source 
motion from rectilinear motion caused by the parallactic 
motion of the Earth around the Sun \citep{gould00}.  The 
microlens parallax is defined by $\pi_{\rm E}={\rm AU}/
\tilde{r}_{\rm E}$, i.e., the ratio of the radius of the 
Earth's orbit around the Sun to the Einstein radius projected
on the observer plane, $\tilde{r}_{\rm E}$.   The subscripts
``$N$'' and ``$E$'' refer to the components of the parallax 
vector projected on the sky in the north and east celestial 
coordinates.

To determine the model parameters, we conduct a massive search 
for solutions over a broad range of mass ratios and separations.  
For this, we hold the binary parameters $s$, $q$, and $\alpha$ 
fixed at a grid of values, while the remaining parameters 
($t_{\rm E}$, $u_0$, $t_0$, and $\rho_\star$) are allowed to 
vary so that the model light curve results in minimum $\chi^2$ 
at each grid point.  We use a Markov Chain Monte Carlo method 
for the $\chi^2$ minimization.  Once the $\chi^2$ minima of the 
individual grid points are determined, the best-fit model is 
obtained by comparing the $\chi^2$ values.  Although a brute search 
in the space of the binary parameters requires a large computation 
time, it is needed to investigate the existence of possible 
degenerate solutions.

For the computation of lensing magnification including the 
finite-source effect, we use the ``magnification map'' technique 
\citep{dong06, dong09}.  In this technique, a magnification map 
is  constructed for a given set of $s$ and $q$ by using the 
ray-shooting method.  Based on this map, the lensing magnification 
corresponding to a location of a source star on the map is computed 
by comparing the number density of rays within the source radius 
with the density on the lens plane.  On the basis of this scheme, 
the computation is accelerated in several ways.  First, the 
algorithm is designed to save computation time by limiting the 
range of ray shooting on the image (lens) plane to a narrow 
annulus encircling the Einstein ring of the primary lens. This 
is based on the fact that only the rays from this region arrive 
at the central perturbation region on the source plane.  Second, 
the algorithm keeps the information of the positions of all the 
rays arriving at the target in the buffer memory space of a 
computer so that the information can be promptly used for fast 
computation of the magnification.  Third, finite-source 
magnification calculation based on ray-shooting is used only 
in the region near the caustic and a simple hexadecapole 
approximation \citep{pejcha09, gould08} is used in other part 
of light curves.

In Figure~\ref{fig:one} -- Figure~\ref{fig:three}, we present the 
model light curves superposed on the data of the individual events.  
The geometry of the source trajectory with respect to the caustic 
location for each event is presented in the inset of each figure.  
The coordinates of the inset are centered at the {\it center of 
magnification} around which the single-lens magnification best 
describes that of the binary lensing. The abscissa and ordinate 
are parallel with and normal to the binary axis, respectively.  
In Table~\ref{table:one}, we also present the values of the 
individual lensing parameters.\footnote{For the event 
MOA-2007-BLG-199/OGLE-2007-BLG-419, we present two sets of solutions 
resulting from the known ``$+u_0\leftrightarrow -u_0$'' degeneracy 
in parallax determination \citep{smith03}.  We find that there exists 
no ``jerk-parallax'' degeneracy, which is another known degeneracy in 
parallax determination \citep{gould04, park04}.} We note that modeling 
the light curve of the event MOA-2007-BLG-199/OGLE-2007-BLG-419 with 
the basic binary parameters does not yield a satisfactory model and 
requires inclusion of additional parallax parameters.  As a result, the 
trajectory of the event is curved.    It is found that the determined 
mass ratios of the three lenses range from 0.24 to 1.0, implying 
that the perturbations are produced by binary companions not planets.  
It is also found that the short strong perturbations are caused by the 
crossing the tip of the caustic.

We note that including the parameters of higher-order effect, 
$\rho_\star$ and $\pivec_{\rm E}$, is necessary to obtain a good 
fit to the data.  But the data reductions, which comes directly 
from the pipelines, might not be adequate to measure these subtle 
parameters accurately.  Since these parameters are irrelevant to 
the main scientific purpose of the paper, and since refined data 
reductions are typically very time-consuming, we present these 
higher-order parameters only for completeness.  


\section{Binary Features}

The light curves of the events we analyzed are mainly characterized 
by the perturbation near the peak similar to the central perturbation 
caused by a planet. However, close examination of the perturbations 
reveals several features that convey the binary nature of the 
perturbation.  In this section, we discuss these features.

The first feature that gives a clue as to the nature of the events
is the strength of the main perturbation.  For all three events, 
the deviations from the smooth base of the light curve exceed  
3 magnitudes.  This strength suggests that the perturbations are 
caused by caustic crossings.

The second feature is the duration of the perturbation.  For all 
events, the duration of the main perturbation is very short compared 
to their time scales.  The short duration of the perturbation can 
be explained either by the caustic being tiny or by the source 
trajectory crossing the tip of a cuspy part of a caustic with a 
non-negligible size.

The third feature is the base magnification just before and after 
the perturbation.  The base magnification of a caustic-crossing 
event is important for the characterization of the lens system 
because it gives a strong constraint on the size of the caustic.  
For the event OGLE-2007-BLG-355/MOA-2007-BLG-278, the base 
magnification is only $A\sim 2$ (corresponding to the normalized 
lens-source separation of $u\sim 0.5$). This implies that the 
caustic size is half as large as the Einstein radius and thus 
the binary origin of the perturbation can be immediately noticed.  
The events OGLE-2007-BLG-137/MOA-2007-BLG-091 and
MOA-2007-BLG-199/OGLE-2007-BLG-419 have moderate base 
magnifications of $A\sim 10$ (corresponding to $u\sim 0.1$). 
Considering that the size of the central caustic produced by a 
giant planet with a mass ratio $q\sim 10^{-3}$ located around 
the Einstein ring of the primary with a separation $s\sim 1.3$ 
is $\Delta\xi_{\rm c}\sim 0.01$, the chance that the main 
perturbations of these events are caused by planets is not high. 
However, one still cannot rule out the possibility of a big 
central caustic produced by a planet located very close to the 
Einstein ring (resonant planet).

\begin{figure*}[t]
\epsscale{0.8}
\plotone{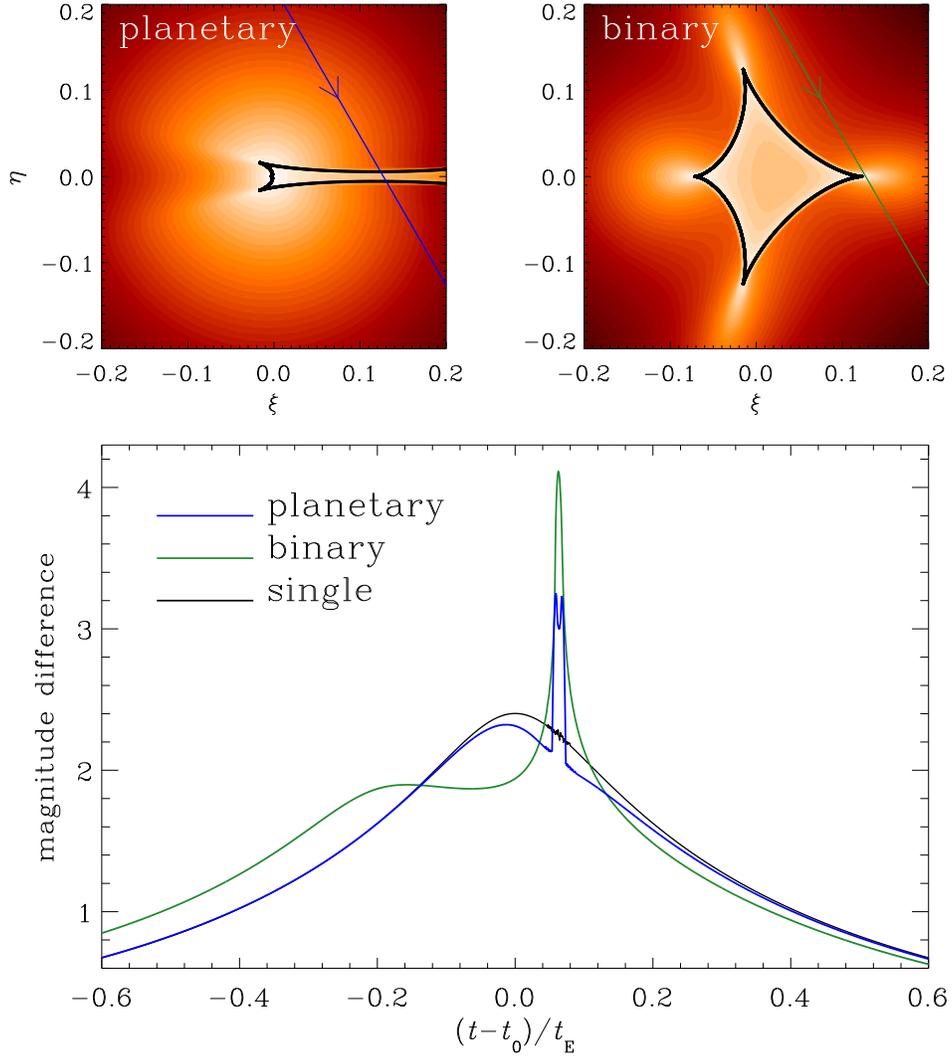}
\caption{\label{fig:four}
The magnification maps of a resonant planetary system (upper 
left panel) and a close binary system (upper right panel) in
the regions around the central caustics of the individual 
systems.  The lower panel shows the light curves resulting 
from the source trajectories marked in the upper panels.
}\end{figure*}

The last feature is the additional secondary perturbation.  The 
central caustic of a resonant planet can become large and thus 
planetary perturbations can occur at a moderate base magnification.
Figure~\ref{fig:four} shows the magnification maps of a resonant 
planetary system in the region around the central caustic.  To 
produce a planetary perturbation at a low magnification, the source 
trajectory should pass close to the cusp corresponding to the tip 
of the arrow-head shaped planetary central caustic such as the one 
shown in Figure~\ref{fig:four}.  In this case, it is difficult to 
produce an additional secondary perturbation because all the other 
cusps are remotely located on the opposite side of the caustic close 
to the primary.  On the other hand, the central caustic produced by 
a binary companion has a diamond shape and thus the geometric chance 
to produce a secondary perturbation by passing an adjacent cusp of 
the caustic is high as demonstrated in Figure~\ref{fig:four}.  
Therefore, the existence of secondary perturbation strongly advocates 
the binary interpretation of the events.

In summary, central perturbations that are characterized by short 
time-scale caustic-crossing features occurring at a low or a 
moderate base magnification with an additional secondary perturbation 
are caused by binary companions.  Considering that three events 
were detected during a single observation season, events with such 
a perturbation feature are common and the feature can be used as a 
diagnostic for the discrimination between the binary and planetary 
interpretations of perturbations.

\section{Conclusion}

We analyzed the light curves of three microlensing events with 
short-term perturbations near the peaks of the light curves. 
From detailed modeling of the light curves, we found that the 
perturbations of the events are caused by binary companions rather 
than planets. From close examination of the light curves combined 
with the underlying physical geometry of the lens system obtained 
from modeling, we found that the short time-scale caustic-crossing 
feature occurring at a low or a moderate base magnification with an 
additional secondary perturbation is a typical feature of binary-lens 
events and thus can be used for the discrimination between the binary 
and planetary interpretations.

\acknowledgments 
We acknowledge the following support: National Research Foundation of 
Korea 2009-0081561 (CH);
NSF AST-0757888 (AG,SD); NASA NNG04GL51G (DD, AG, RP); Polish MNiSW 
N20303032/4275 (AU); HST-GO-11311 (KS); NSF AST-0708890, NASA NNX07AL71G 
(DPB); MEXT19015005, JSPS18749004 (TS); Marden Fund of NZ (IAB, JBH, DJS, 
SLS, PCMY); Foundation for Research Science and Technology of NZ (IAB, LS);
Korea Astronomy \& Space Science Institute (B-GP, C-UL); Deutsche 
Forschungsgemeinschaft (CSB); PPARC, EU FP6 programme ``ANGLES'' (LW, NJR); 
PPARC (RoboNet); Dill Faulkes Educational Trust (Faulkes Telescope North); 
Grants JSPS18253002, JSPS20340052, and JSPS19340058 (MOA); Marsden Fund of 
NZ (IAB, PCMY); Foundation for Research Science and Technology of NZ.
The operation of Mt.~Canopus Observatory is supported in part by the 
financial contribution from David Warren.

\include{table1}

\end{document}